\documentclass[12pt,a4paper]{narms}

\usepackage{subfigure} 
\usepackage{psfig}
\usepackage{timesmt}   
\usepackage{chicaco}   
\usepackage{amsmath,amsfonts}   
\usepackage{graphicx}

\makeatletter
\renewcommand{\section}{\@startsection%
{section}%
{1}%
{0mm}%
{- \baselineskip}%
{0.15\baselineskip}%
{\normalfont\normalsize}}%

\renewcommand{\subsection}{\@startsection
{subsection}%
{2}%
{0mm}%
{-\baselineskip}%
{0.15\baselineskip}%
{\normalfont\normalsize}}%
\makeatother

\begin{document}
\addtocounter{page}{118}
\title{{\small \it \flushright Reliability and Optimization of Structural Systems - Straub (ed)\\
c2010 Taylor \& Francis Group, London, ISBN 978-0-415-88179-1\\}
\vspace{5mm}
Identification of the parameters of complex constitutive models:\\ Least squares minimization vs. Bayesian updating}
\author{\large {Thomas Most}\\
{\em Research Training Group 1462, Bauhaus-Universit\"at Weimar, Germany}\\
}
\date{}

\abstract{In this study the common least-squares minimization approach 
is compared to the Bayesian updating procedure. In the content of material parameter identification 
the posterior parameter density function is obtained from its prior and the likelihood function of the measurements.
By using Markov Chain Monte Carlo methods, such as the Metropolis-Hastings algorithm \cite{Hastings1970}, 
the global density function including local peaks can be computed.
Thus this procedure enables an accurate evaluation of the global parameter quality.
However, the computational effort is remarkable larger compared to the minimization approach.
Thus several methodologies for an efficient approximation
of the likelihood function are discussed in the present study.}


\maketitle
\frenchspacing   


\section{INTRODUCTION}
In all fields of computational mechanics numerical models are used to analyze
engineering problems.
In material modeling many types of constitutive 
formulations exist for a wide range of materials.
Generally the parameters of these models are identified
from experimental data.
Beside manual identification by try and error or based on experiences
fully automatic identification procedures using optimization strategies have become
popular in the last years.
Nevertheless, not only the parameter values itself are of interest for a later
numerical analysis, but also their accuracy which is mainly influenced by the measurement errors
and the type of experiments. For very complex constitutive models it is very difficult to judge
in advance which experiments are required to enable a sufficient identifiability
of all parameters.
Therefor an estimation of the parameter uncertainties is very useful.
Based on the determined optimal parameter set as the result of a least squares
minimization, where the gap between experimental and numerical response data is minimized,
the parameter covariances can be estimated based on the measurement errors and a
linearization of the model \shortcite{Ledesma1996}.
This approach was derived for a convex optimization problem.
Applications for non-convex problems using complex material models in geomechanical applications
can be found in \shortcite{Hofmann_2009_EUROTUN}, \shortcite{Knabe_2009_EUROTUN}.

Alternatively to the ordinary least squares formulation
Kalman filter identification has been introduced \shortcite{Cividini1983}, \shortcite{Bittani1984}.
This technique is a Bayesian estimator using prior information
to identify the parameters. Similar to the covariance analysis in least squares minimization
the model itself is linearized and the parameter distributions are taken as Gaussian which implies a
convex optimization problem.

Another method uses Bayesian inference to estimate the parameters including their statistical properties
\shortcite{Beck1977}.
This approach can be applied for  non-convex problems and arbitrary parameter and measurement error distributions
and works without any linearization of the numerical model.

Recently inverse approximation schemes using neural networks have been proposed to identify the constitutive parameters 
 \shortcite{Novak2006},\shortcite{Most_2007_AICC}.
In \shortcite{Unger2009h} this approach has been extended using Bayesian neural networks in order
to estimate the parameter accuracy. In opposite to the other procedures there measurement errors
are neglected and the parameter uncertainty arises only from the identification procedure.
This results in a vanishing uncertainty if the number of training samples is increased dramatically.
Therefor this procedure is not discussed further in this study.

\section{LEAST SQUARES MINIMIZATION}
\subsection{Maximum Likelihood formulation}

A deterministic models which relates a set of responses
$\mathbf{x}=\mathbf{x}(\mathbf{p})$ with the model parameters
$\mathbf{p}$ is assumed to be given. 
In the parameter identification procedure the measurements $\mathbf{x^*}=[x^*_1\ldots
x^*_m]^T$ are obtained by experiments.

The likelihood of the parameters is proportional to the conditional probability 
of measurements $\mathbf{x^*}$ from a given parameter set $\mathbf{p}$ \shortcite{Ledesma1996}: 
\begin{equation}
L=k\cdot f(\mathbf{x^*}| \mathbf{p}).
\end{equation} 
If the chosen model is correct the gap between numerical responses and measurements
$(\mathbf{x^*} - \mathbf{x})$ is caused only by the measurement errors.
Thus $P(\mathbf{x^*}| \mathbf{p})$ is equivalent to the probability of reproducing the measurement errors.
Assuming a multivariate normal distribution we obtain
\begin{equation}
	P(\mathbf{x^*} - \mathbf{x}) = \frac{1}{\sqrt{(2\pi)^m |\mathbf{C}_\mathbf{xx}|}}
	\exp\left[-\frac{1}{2}(\mathbf{x^*} - \mathbf{x})^T\mathbf{C}_\mathbf{xx}^{-1}(\mathbf{x^*} - \mathbf{x})\right].
\label{likeli}
\end{equation} 
Maximizing the likelihood $L$ is equivalent to minimize $S=-2\ln L$.
This yields to the well known least squares objective function:
\begin{equation}
	J=\left(\mathbf{x^*} - \mathbf{x}\right)^T\mathbf{C}_\mathbf{xx}^{-1}\left(\mathbf{x^*} - \mathbf{x}\right),
\label{objective}
\end{equation}
where $\mathbf{C}_\mathbf{xx}$	is the covariance matrix of the measurements.
In \shortcite{Ledesma1996} it is shown, that the optimal weights used in the objective function
are 
\begin{equation}
\mathbf{W} = \mathbf{C}_\mathbf{xx}^{-1}.
\end{equation}
If the objective function is linearized, the following update scheme is obtained
\begin{equation}
  	\Delta \mathbf{p} = \left(\mathbf{A}^T\mathbf{C}_\mathbf{xx}^{-1}\mathbf{A}\right)^{-1}\mathbf{A}^T\mathbf{C}_\mathbf{xx}^{-1} \Delta \mathbf{x},
\label{lin_model}
\end{equation} 
where $\mathbf{A}$ is the sensitivity matrix of the parameters with respect to the model responses 
\begin{equation}
	\mathbf{A}=\frac{\partial \mathbf{x}}{\partial \mathbf{p}}.
\end{equation} 
The update scheme can be used to obtain the optimal parameter set with the minimum
objective function value. This approach is a gradient based method which leads to the global minimum without fail 
only for a convex optimization problem. If this method is applied for a non-convex problem,
it may stuck in a local minimum. This is the case for more complex constitutive models.
For this reason often global optimization schemes are used to identify the optimal parameter set.
One of these methods is presented in the next section.

If the optimal parameter set is finally identified,
the covariances of the parameters $\mathbf{C}_\mathbf{pp}$ can be estimated by using the linearized relation from Eq.(\ref{lin_model})
which leads to the so-called Markov estimator
\begin{equation}
	\mathbf{C}_\mathbf{pp} = \left(\mathbf{A}_{opt}^T \mathbf{C}_\mathbf{xx}^{-1}\mathbf{A}_{opt}\right)^{-1}.
\end{equation}
In \shortcite{Hofmann_2009_EUROTUN} and \shortcite{Knabe_2009_EUROTUN}
this estimator is applied also for non-convex optimization problem, which means that
around the global optimum a local convex problem is assumed and the 
model is linearized at this optimum. Both, parameter and measurement distributions
are assumed to be Gaussian within this approach.

\subsection{Particle swarm optimization}
In our study a population based global optimization algorithm, the particle swarm optimization (PSO), is applied 
which enables detecting the global optimum even if several local minima exist.
Each population consists of a given number of  particles
where each particle
position is equivalent to a set of parameters $\mathbf{p}_i^k$ and 
is updated in each iteration step $k$ by the simple scheme \shortcite{Kennedy1995}
\begin{equation}
\begin{aligned}
\mathbf{p}_i^{k+1} &= \mathbf{p}_i^k + \mathbf{v}_i^{k+1}\\
\mathbf{v}_i^{k+1} &= \omega \mathbf{v}_i^k + c_1 \mathbf{r}_1 (\mathbf{P}_i^k - \mathbf{p}_i^k) 
+ c_2 \mathbf{r}_2 (\mathbf{P}_g^k -\mathbf{p}_i^k)\\
\end{aligned}
\end{equation}
where $c_1$ and $c_2$ are constants and $\mathbf{r}_1$ and $\mathbf{r}_2$ are vectors of uniformly
distributed random variables between zero and one.
$\mathbf{P}_i^k$ is the best position of a single particle and $\mathbf{P}_g^k$ is the global best position.
We extend the original concept with passive congregation \shortcite{He2004}
\begin{equation}
\mathbf{v}_i^{k+1} = \omega \mathbf{v}_i^k + c_1 \mathbf{r}_1 (\mathbf{P}_i^k - \mathbf{p}_i^k)
+ c_2 \mathbf{r}_2 (\mathbf{P}_g^k -\mathbf{p}_i^k)
 + c_3 \mathbf{r}_3 (\mathbf{R}_i^k -\mathbf{p}_i^k)
\end{equation}
where an additional term is added to the original scheme 
to decrease the risk of running into a local minimum.
This term $\mathbf{R}_i^k$ is defined as the best position of a randomly chosen particle.
The required constants $c_1$, $c_2$ and $c_3$ are given in \shortcite{He2004}.
If new particle positions are outside the parameter boundaries they are
 corrected by a harmony search scheme according to \shortcite{Li2007}.

\section{BAYESIAN ESTIMATORS}
\subsection{Kalman filter}
Kalman filter technology was firstly applied for parameter identification
in \shortcite{Cividini1983}, \shortcite{Bittani1984}.
More recent applications can be found in \shortcite{Bolzon2002}, \shortcite{Fedele2006}, \shortcite{Furukawa2009}.
In this approach the measurement noise is assumed to be a time-dependent white Gaussian random process
characterized by zero mean, zero cross-correlation, and a time-dependent covariance matrix
\begin{equation}
\begin{aligned}
E(\boldsymbol{ \epsilon}^{(t)}) &=\mathbf{0}, \\
E(\boldsymbol{ \epsilon}^{(t)}\boldsymbol{ \epsilon}^{(s)^T}) &= \mathbf{0}, s \neq t\\
E(\boldsymbol{ \epsilon}^{(t)}\boldsymbol{ \epsilon}^{(t)^T}) &= \mathbf{C}_\mathbf{xx}^{(t)}.
\end{aligned}
\end{equation}
Using a linearization of the model the following update scheme can be derived
\begin{equation}
\begin{aligned}
\mathbf{\hat p}^{(t)} &= \mathbf{\hat p}^{(t-1)} + \mathbf{K}^{(t)}(\mathbf{x^*} - \mathbf{x}(\mathbf{\hat p}^{(t-1)}))\\
\mathbf{\hat C}_\mathbf{pp}^{(t)} &= \mathbf{\hat C}_\mathbf{pp}^{(t-1)} - \mathbf{K}^{(t)}\mathbf{A}^{(t)}\mathbf{\hat C}_\mathbf{pp}^{(t-1)}
\end{aligned}
\end{equation}
where $\mathbf{K}$ is the Kalman gain matrix 
\begin{equation}
\mathbf{K}^{(t)} = \mathbf{\hat C}_\mathbf{pp}^{(t-1)} \mathbf{A}^{(t)^T}\left[\mathbf{A}^{(t)}\mathbf{\hat C}_\mathbf{pp}^{(t-1)}
\mathbf{A}^{(t)^T}+\mathbf{C}_\mathbf{xx}^{(t)}\right]^{-1}
\end{equation}
and $\mathbf{A}$ is again the sensitivity matrix.
As initial setting generally  the prior information is taken
\begin{equation}
\mathbf{\hat p}^{(0)} = \mathbf{p}_0, \quad \mathbf{\hat C}_\mathbf{pp}^{(0)}= \mathbf{C}_{\mathbf{pp}_0}.
\end{equation}
Due to the linearization of the model and the assumed Gaussian measurement noise,
the parameter distributions are implicitly assumed to be Gaussian.
In a static approach the time-step $t$ is equivalent to the iteration step and the measurement covariance matrix remains constant.

\shortcite{Cividini1983} has shown, that for vanishing prior information, where $(\mathbf{C}_{\mathbf{pp}_0})^{-1}$
tends to zero, the estimated parameter covariance is equivalent to the Markov estimator.

\subsection{Bayesian updating}
Bayesian updating is based on the well-known Bayes' theorem
which shows the relation between one conditional probability and its inverse
\begin{equation}
    P(A|B) = \frac{P(B | A)\, P(A)}{P(B)}. 
\end{equation}
Assuming a prior multivariate distribution of the parameters $\pi_{prior}(\mathbf{p})$
and a likelihood-function $\pi(\mathbf{x^*}|\mathbf{p})$ for the distribution of the measurements
the conditional probability of the parameters can be obtained as \shortcite{Beck1977}
\begin{equation}
  	\pi_{posterior}(\mathbf{p})=\pi(\mathbf{p}|\mathbf{x^*}) = \frac{\pi_{prior}(\mathbf{p}) \cdot \pi(\mathbf{x^*}|\mathbf{p})}{\pi(\mathbf{x^*})}.
\end{equation}
Since the normalization constant $\pi(\mathbf{x^*})$ is difficult to 
determine due to the only implicitly given distribution function,
realizations of $\pi_{posterior}(\mathbf{p})$ are obtained by Markov-Chain Monte Carlo Simulation \shortcite{Hastings1970}.
For jointly Gaussian measurement errors the likelihood function is equivalent to Eq.(\ref{likeli}).

\subsection{Metropolis-Hastings algorithm}

The Metropolis-Hastings algorithm \shortcite{Hastings1970}
enables the sampling of an implicitly given not necessarily scaled density function. Therefor
it is very important in interferential problems.
This method is a Monte Carlo Simulation using a first order Markov chain,
where the next sample depends only on the current state.

The algorithm generates a sequence of samples $\{\mathbf{p}^{(1)}, \mathbf{p}^{(2)},\ldots,  \mathbf{p}^{(n)}\}$
which statistically converge to the given distribution, which is in our case the Bayesian posterior
multivariate parameter distribution. The algorithm works as follows:
\begin{enumerate}
\item Sample candidate $\mathbf{p^*}$ from jumping distribution $q(\mathbf{p^*},\mathbf{p}^{(t-1)})$
\item Calculate 
		\begin{equation}
		\alpha = \min \left[ 1, \frac{\pi(\mathbf{p^*}|\mathbf{x^*})}{\pi(\mathbf{p}^{(t-1)}|\mathbf{x^*})}
		\frac{q(\mathbf{p}^{(t-1)},\mathbf{p^*})}{q(\mathbf{p^*},\mathbf{p}^{(t-1)})}
		\right]
		\end{equation}
\item Sample uniformly distributed $U\in (0.0; 1.0)$
\item If $U \leq \alpha$ accept $\mathbf{p}^{(t)} = \mathbf{p^*}$, otherwise $\mathbf{p}^{(t)} = \mathbf{p}^{(t-1)}$
\item Return to step 1
\end{enumerate}
The jumping distribution is taken generally to be a Gaussian or uniform distribution. If it is symmetric\\
$q(\mathbf{p}^{(t-1)},\mathbf{p^*})/q(\mathbf{p^*},\mathbf{p}^{(t-1)})$ is equal to one.
The initial parameter set is chosen randomly in the given parameter ranges.
In this study the prior parameter distribution is taken as a uniform distribution
in these parameter ranges. Thus the Metropolis-Hastings algorithm will only accept
samples inside this boundaries.
At the beginning the generated sequence is biased depending on the starting point and the jumping distribution.
But after a certain burn-in phase it turns into an ergodic process.
Thus the statistical evaluation of the parameter distributions is done be neglecting a given number of the first samples.

The variances of the jumping distribution strongly influences acceptance rate (AR) and convergence speed.
If the distribution variances are taken too small, the convergence is very low and almost all samples will be accepted.
An optimal statistical evaluation is enabled if the acceptance rate is between 10 and 30 \% as shown in \shortcite{Unger2009}.

\begin{figure}[th]
\vspace{-5mm}
\centering
\begin{minipage}[c]{0.48\textwidth}
\hspace{-5mm}\includegraphics[width=1.05\textwidth]{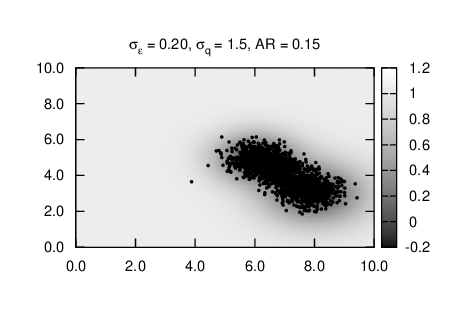}
\end{minipage}\\
\vspace{-12mm}
\begin{minipage}[c]{0.48\textwidth}
\hspace{-5mm}\includegraphics[width=1.05\textwidth]{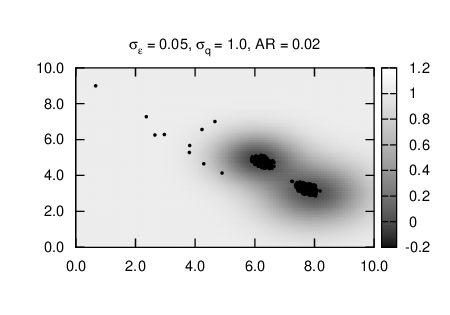}
\end{minipage}
\begin{minipage}[c]{0.48\textwidth}
\hspace{-5mm}\includegraphics[width=1.05\textwidth]{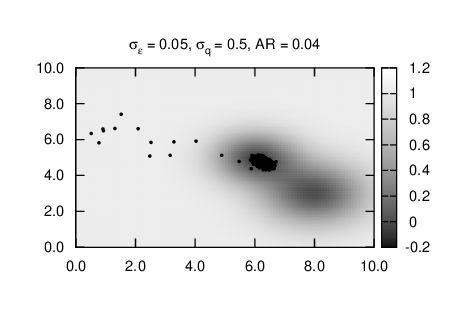}
\end{minipage}
\caption{Samples from Metropolis-Hastings algorithm for a non-convex problem with different measurement errors}
\label{nonconvex_acceptance}
\end{figure}

For non-convex problems the variances of the jumping distribution have to be chosen
in that way, that the peaks of the likelihood function are sufficiently covered.
This is shown in Figure \ref{nonconvex_acceptance} and Table \ref{nonconvex_results} depending on the measurement errors.
If the measurement errors are small and the variances of the jumping distribution are to small,
 only one local optimum is covered by the sampling procedure.
But if the variances are chosen large enough the acceptance rate is very small and a large number of samples
is necessary to obtain statistically accurate results.

\begin{table}[th]
\centering
\small
\begin{tabular}{cccccc}
    \hline
	\multicolumn{6}{c}{Bayes' updating} \rule[0mm]{0pt}{2.5ex}\\
	$\sigma_\epsilon$  & $\sigma_q$ & $AR$ & No. samples & $\tilde \sigma_{X_1}$ & $\tilde \sigma_{X_2}$\rule[-2mm]{0pt}{2ex}\\
	
   \hline
    0.05	& 1.0 & 0.02 &100000& 0.81 & 0.80\rule[0mm]{0pt}{2.5ex}\\
	0.10	& 1.5 & 0.05 &44000& 0.77 & 0.76\\ 
	0.20	& 1.5 & 0.15 &13000& 0.79 & 0.81\\ 

    0.05	& 0.5 & 0.04 &44000& 0.13 & 0.12\rule[0mm]{0pt}{2.5ex}\\
	0.10	& 0.5 & 0.20 &10000& 0.78 & 0.78\\ 
	0.20	& 0.5 & 0.50 &4000& 0.79 & 0.79\\ 
   \hline
  \end{tabular}
\caption{Estimated parameter variation for a non-convex problem with different measurement errors and variances of the jumping distribution}
\label{nonconvex_results}
\end{table}

\subsection{Scaling of measurement errors}

In the previous section it was mentioned, that for small measurement errors a large number of samples may be required
if the optimization problem is non-convex.
For this reason two possibilities to reduce this numerical effort are discussed.

The first idea is to scale the covariances of the measurement errors by a constant factor $a$
\begin{equation}
\mathbf{C}_\mathbf{xx}^{scaled} = a \cdot \mathbf{C}_\mathbf{xx}^{orig}.
\end{equation}
Based on the scaled likelihood function a sequence of samples is computed with the Metropolis-Hastings algorithm.
The estimates of the parameter mean values and covariances of the original distribution can be computed as follows
\begin{equation}
\begin{aligned}
	\bar p_k &= \frac{1}{n} \sum_{i=1}^{n} p^{(i)}_k \frac{f_{orig}(\mathbf{p}^{(i)})}{f_{scaled}(\mathbf{p}^{(i)})}\\
	C_{kl} &= \frac{1}{n-1} \sum_{i=1}^{n} (p^{(i)}_k - \bar p_k)(p^{(i)}_l - \bar p_l) 
	\frac{f_{orig}(\mathbf{p}^{(i)})}{f_{scaled}(\mathbf{p}^{(i)})}
\end{aligned}
\end{equation}
where the joint densities functions $f_{orig}$ and $f_{scaled}$
have to be known. Both are equivalent to the corresponding likelihood function scaled by the normalization constant.
Since this normalization constant is not known it has to be estimated from the samples.
In our opinion this is only possible by assuming a certain distribution type.
This would suspend the advances of the Bayesian updating approach and would limit the method again to convex optimization problems.
Therefor this scaling procedure is not investigated further in this study.

\subsection{Approximation of the likelihood function}

Another approach to reduce the numerical effort is to approximate the likelihood function
by a suitable meta-model and using this approximation function instead of real model
solutions.
In \shortcite{Orlande2008} this was shown by using radial basis function approximation.
\begin{figure}[th]
\vspace{-15mm}
\begin{minipage}[c]{0.48\textwidth}
\hspace{-7mm}\includegraphics[width=1.1\textwidth]{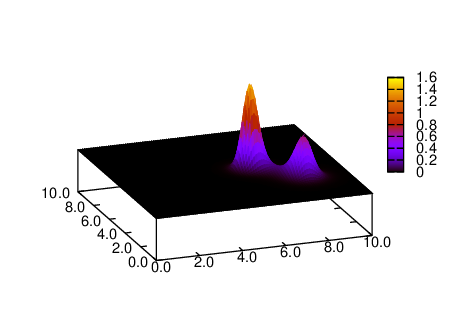}

\vspace{-8mm}
\hspace{17mm}$\exp\left[-\frac{1}{2}(\mathbf{x^*} - \mathbf{x})^T\mathbf{C}_\mathbf{xx}^{-1}(\mathbf{x^*} - \mathbf{x})\right]$
\end{minipage}
\begin{minipage}[c]{0.48\textwidth}
\hspace{-7mm}\includegraphics[width=1.1\textwidth]{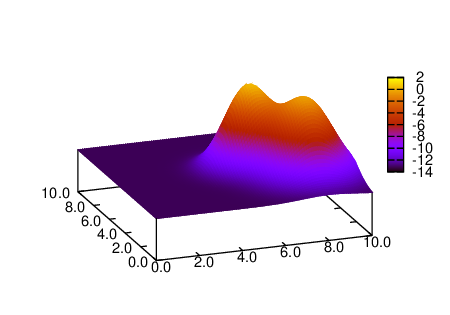}

\vspace{-8mm}
\hspace{20mm}$-\frac{1}{2}(\mathbf{x^*} - \mathbf{x})^T\mathbf{C}_\mathbf{xx}^{-1}(\mathbf{x^*} - \mathbf{x})$
\end{minipage}

\vspace{3mm}
\caption{Likelihood function with sharp peaks and corresponding exponential term for small measurement errors}
\label{nonconvex_approximation}
\end{figure}
Dependent on the variance of the measurement noise
the likelihood function contains sharp peaks around the local optima as shown in Figure \ref{nonconvex_approximation}.
This requires an increased number of support points in the meta-model approach.
For this reason in this study only the exponential term is approximated, which is equivalent
to the objective function in Eq.(\ref{objective}). Then the measurement noise variance is only a scaling factor with no influence on the
approximation quality.
As meta-model approach interpolating Moving Least Squares \shortcite{Most_2008_EABE} is applied,
since it can handle clustered support points and high gradients in the approximation function. 

\begin{figure}[t]
\vspace{-5mm}
\begin{minipage}[c]{0.48\textwidth}
\hspace{-5mm}\includegraphics[width=1.05\textwidth]{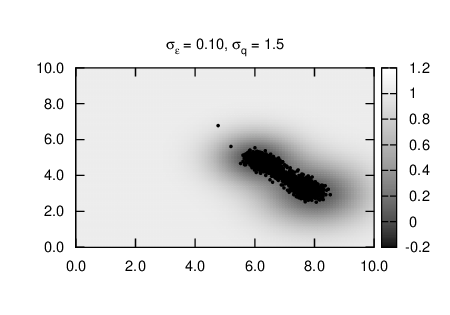}
\end{minipage}
\begin{minipage}[c]{0.48\textwidth}
\hspace{-5mm}\includegraphics[width=1.05\textwidth]{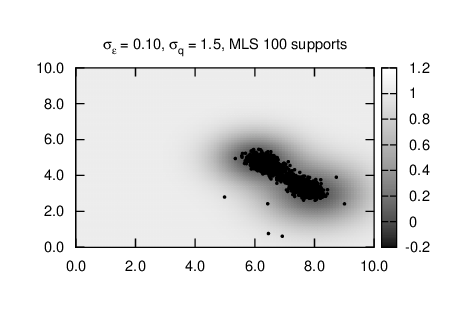}
\end{minipage}
\caption{Computed parameter samples by using the original and approximated likelihood function}
\label{nonconvex_mls}
\end{figure}

In Figure \ref{nonconvex_mls} and Table \ref{nonconvex_mls_results}
the results with original and approximated likelihood function are compared for the non-convex problem.
The results indicate a very good agreement in the parameter estimates. The numerical effort is reduced dramatically,
since only 100 evaluations of the real model are required instead of 44000 by using the original likelihood function.

\begin{table}[th]
\centering
\small
\begin{tabular}{lcccc}
    \hline
	\multicolumn{5}{c}{MLS approximation $\sigma_\epsilon = 0.10$, $\sigma_q = 1.5$} \rule[0mm]{0pt}{2.5ex}\\
	&  $\tilde \mu_{X_1}$ & $\tilde \mu_{X_2}$ & $\tilde \sigma_{X_1}$ & $\tilde \sigma_{X_2}$\rule[-2mm]{0pt}{2ex}\\
	
   \hline
	Original likelihood & 7.22 & 3.83 & 0.77 & 0.76\rule[0mm]{0pt}{2.5ex}\\ 
	MLS 100 supports 	& 7.23 & 3.98 & 0.71 & 0.70 \\ 
	MLS 200 supports 	& 6.91 & 4.08 & 0.78 & 0.73 \\ 
	MLS 500 supports 	& 7.05 & 3.95 & 0.73 & 0.73 \\ 
   \hline
  \end{tabular}
\caption{Estimated parameters by using the original and approximated likelihood function}
\label{nonconvex_mls_results}
\end{table}

\section{NUMERICAL EXAMPLES}
\subsection{1D elasto-plastic model without hardening}

\begin{figure}[p]
\centering
\begin{minipage}[c]{0.48\textwidth}
\includegraphics[width=\textwidth]{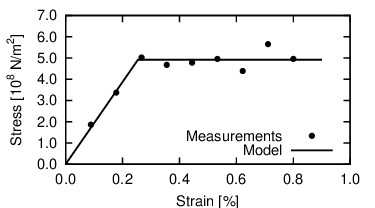}
\end{minipage}
\caption{Stress-strain dependence of the 1D elasto-plastic model}
\label{convex_measurements}
\centering
\begin{minipage}[c]{0.48\textwidth}
\hspace{-2mm}\includegraphics[width=1.05\textwidth]{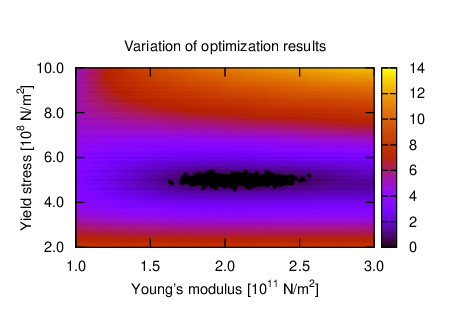}
\end{minipage}\\

\vspace{-8mm}
\begin{minipage}[c]{0.48\textwidth}
\hspace{-2mm}\includegraphics[width=1.05\textwidth]{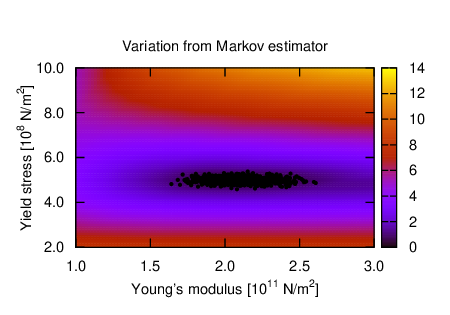}
\end{minipage}
\begin{minipage}[c]{0.48\textwidth}
\hspace{-2mm}\includegraphics[width=1.05\textwidth]{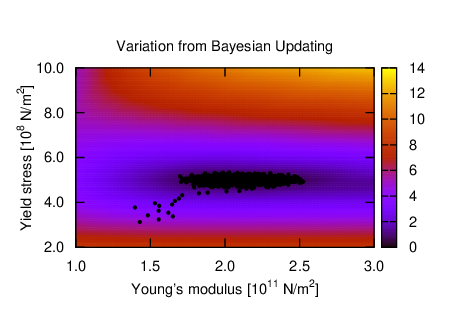}
\end{minipage}
\caption{Deterministic objective function of the 1D elasto-plastic model 
including the parameter samples from the different approaches for $\sigma_\epsilon = 0.2\cdot 10^{8} \text{N/m}^2$}
\label{convex_samples}
\end{figure}
In this analytical example an one-dimensional elasto-plastic model with two parameters, the Young's modulus $E$ and the yield
stress $\sigma_{Y}$ is investigated.
As reference values $E_{ref} = 2.1 \cdot 10^{11} \text{N/m}^2$ and $\sigma_{Y,ref}= 5.0 \cdot 10^8 \text{N/m}^2$
are chosen.
By applying constant Gaussian measurement noise $\sigma_\epsilon$ for all observations
1000 samples are generated and for each sample the optimal parameter set is obtained by a gradient-based optimization
method.
The resulting variations of the identified parameters are shown in Table \ref{convex_results}.

Additionally the Markov estimator is used at the parameter reference values and the Kalman filter
approach is applied by using a large prior covariance $\mathbf{C}_\mathbf{pp}^0 = diag[100,100] \cdot 10^8 \text{N/m}^2$
in order to reduce the effect of the prior information.
The results of both methods coincide excellent with the sample values due to the convexity of the problem.
In Figure \ref{convex_samples} the resulting parameter samples are shown.
\begin{table}[th]
\centering
\small
\begin{tabular}{lllcccc}
    \hline
	&&&\multicolumn{3}{c}{$\sigma_\epsilon [10^{8} \text{N/m}^2]$}\rule[0mm]{0pt}{2.5ex}\\
	&&& $0.05$ & $0.10$ & $0.20$\\
	
    \hline
	\multicolumn{1}{l}{Sample variation}
	&$\sigma_E$ & $[10^{11} \text{N/m}^2]$ 				& 0.025 & 0.052 & 0.157\rule[0mm]{0pt}{2.5ex}\\
	&$\sigma_{\sigma_{Y}}$ & $[10^{8} \text{N/m}^2]$ 	& 0.019 & 0.036 & 0.117\\
	\multicolumn{1}{l}{Markov estimator}
	&$\tilde \sigma_E$ & $[10^{11} \text{N/m}^2]$		& 0.025 & 0.050 & 0.151\rule[0mm]{0pt}{2.5ex}\\
	&$\tilde \sigma_{\sigma_{Y}}$ & $[10^{8} \text{N/m}^2]$ & 0.019 & 0.038 & 0.113\\
	\multicolumn{1}{l}{Kalman filter}
	&$\tilde \sigma_E$ & $[10^{11} \text{N/m}^2]$		& 0.025 & 0.050 & 0.151\rule[0mm]{0pt}{2.5ex}\\
	&$\tilde \sigma_{\sigma_{Y}}$ & $[10^{8} \text{N/m}^2]$ & 0.019 & 0.038 & 0.113\\
	\multicolumn{1}{l}{Bayes' updating}
	&$\tilde \sigma_E$ & $[10^{11} \text{N/m}^2]$		& 0.025 & 0.050 & 0.145\rule[0mm]{0pt}{2.5ex}\\
	&$\tilde \sigma_{\sigma_{Y}}$ & $[10^{8} \text{N/m}^2]$ & 0.019 & 0.038 & 0.115\\
   \hline
\end{tabular}
\caption{Estimated parameter variations for the 1D elasto-plastic model}
\label{convex_results}
\end{table}

Finally the Bayesian updating procedure is used
whereby the prior distributions are chosen to be uniform between the optimization bounds 
 $E \in (1.0;3.0)$, $\sigma_Y \in (2.0; 10.0)$. The resulting
 estimates are given in Table \ref{convex_results}. In Figure \ref{convex_samples}
 the samples from the Metropolis-Hastings algorithm are shown, where the burn-in phase can be seen.
 The first 1000 samples are not considered in the statistical evaluation in order to get unbiased estimates.
The results agree very well with the other approaches. Only for a larger measurement noise minor
deviation can been observed which might be caused by the linearization in the Markov estimator and Kalman filter approaches.

\subsection{Bilinear interface model for concrete cracking}
\begin{figure}[th]
\hspace{5mm}\begin{minipage}[c]{0.48\textwidth}
\centering
\hspace{5mm}\includegraphics[width=0.8\textwidth]{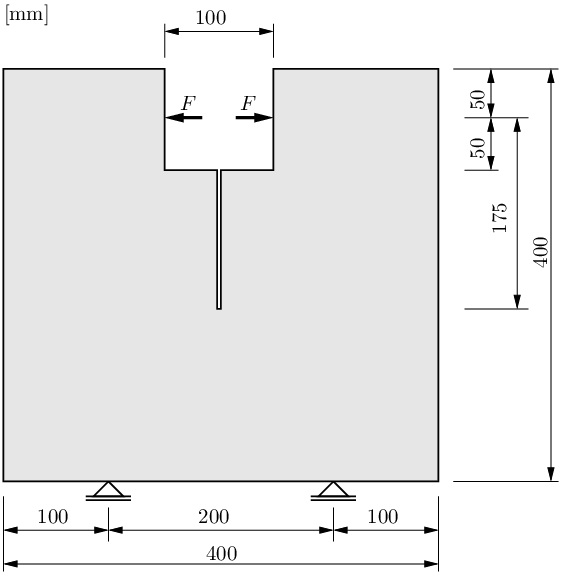}
\end{minipage}
\begin{minipage}[c]{0.48\textwidth}
\centering
\vspace{3mm}
\includegraphics[width=0.8\textwidth]{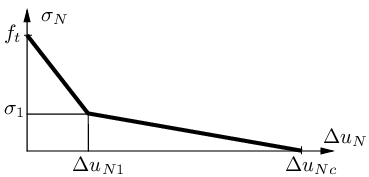}
\end{minipage}
\caption{Wedge splitting test with system geometry and bilinear interface model}
\label{wedge_system}
\end{figure}

\begin{figure}[th]
\hspace{5mm}\includegraphics[width=0.48\textwidth]{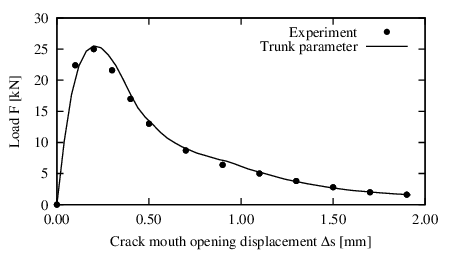}
\includegraphics[width=0.48\textwidth]{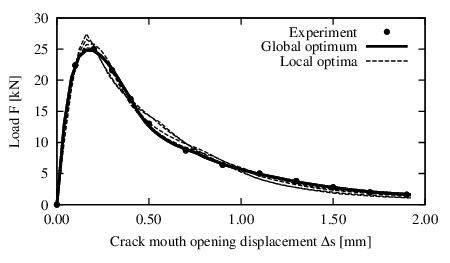}
\caption{Load displacement curves for the wedge splitting test using the identified parameters}
\label{wedge_curves}
\end{figure}

In this example the parameters of a cohesive crack model are identified.
The investigations are based on the experimental program of \shortcite{Trunk1999}.
In Figure \ref{wedge_system} the concrete specimen is shown with the corresponding
dimensions.
The numerical analysis is carried out using a finite element model 
with a predefined crack discretized by finite interface elements.
Further details about the numerical suimulation can be found in
\shortcite{Most_2005_PhD}.
For the interface elements a bilinear softening law is used shown in principle
in Figure \ref{wedge_system}. The remaining base material is assumed to be linear elastic.
The following parameters have been indentified within the experimental program
\begin{itemize}
\item Young's modulus $E =2.83\cdot 10^{10} \mbox{ N/m}^2$
\item Tensile strength $f_t=2.27\cdot 10^{6} \mbox{ N/m}^2$
\item Specific fracture energy $G_f=285 \mbox{ Nm/m}^2$
\item Softening shape parameters $\alpha_\sigma=\sigma_1/f_t=0.163$, $\alpha_u=\Delta u_{N1}/\Delta u_{Nc}=0.242$
\end{itemize}
The corresponding measurements and the numerical load displacement curve is shown in Figure \ref{wedge_curves}

In a first step the measurements are taken as deterministic and 100 optimization runs with random start points 
using a gradient based and the particle swarm
approach are carried out. The results are given in Table \ref{wedge_optim_results}. Due to the existence
of several local minima the gradient based methods does not converge to the global optimum in almost the half of the runs,
whereby the PSO method is successfull in almost all cases .
\begin{table}[t]
\centering
\small
\begin{tabular}{lccccccc}
    \hline
	&&&&\multicolumn{2}{c}{Global}& \multicolumn{2}{c}{Total}\rule[0mm]{0pt}{2.5ex}\\
	&&& Best & Mean & Std. & Mean & Std. \rule[-2mm]{0pt}{2ex}\\
   \hline
   \multicolumn{1}{l}{Gradient (success rate $51 \%$)} 
    &$E$ & $[10^{10} \mbox{ N/m}^2]$   		  &   3.85 & 3.89 & 0.11 & 3.72 & 0.39   \rule[0mm]{0pt}{2.5ex}\\
	&$f_t$ & $[10^{6} \mbox{ N/m}^2]$		  &   2.04 & 2.03 & 0.02 & 2.18 & 0.38  \\
	&$G_f$ & $[10^{2} \mbox{ Nm/m}^2]$		  &   2.84 & 2.84 & 0.01 & 2.81 & 0.07  \\
	&$\alpha_\sigma$ & $[10^{-1}]$			  &   1.57 & 1.58 & 0.05 & 1.86 & 0.51  \\
	&$\alpha_u$ & $[10^{-1}]$				  &   2.54 & 2.55 & 0.07 & 2.46 & 0.41  \\

   \multicolumn{1}{l}{PSO (success rate $89 \%$)} 

    &$E$ & $[10^{10} \mbox{ N/m}^2]$   		  &   3.85 & 3.82 & 0.05 &  3.74 & 0.28  \rule[0mm]{0pt}{2.5ex}\\
	&$f_t$ & $[10^{6} \mbox{ N/m}^2]$		  &   2.04 & 2.05 & 0.01 &  2.13 & 0.31 \\
	&$G_f$ & $[10^{2} \mbox{ Nm/m}^2]$		  &   2.84 & 2.84 & 0.00 &  2.82 & 0.05 \\
	&$\alpha_\sigma$ & $[10^{-1}]$			  &   1.57 & 1.57 & 0.02 &  1.70 & 0.42 \\
	&$\alpha_u$ & $[10^{-1}]$				  &   2.54 & 2.53 & 0.03 &  2.44 & 0.32 \\
   \hline
  \end{tabular}
\caption{Optimization results for the wedge splitting test}
\label{wedge_optim_results}
\end{table}
In  Table \ref{wedge_optim_results} the statistics of the runs ending close to the global optimum
and of all runs are compared. The results indicate a very good accuracy of the PSO method which is even better
than the gradient based approach.

\begin{figure}[th]
\centering
\includegraphics[width=0.43\textwidth]{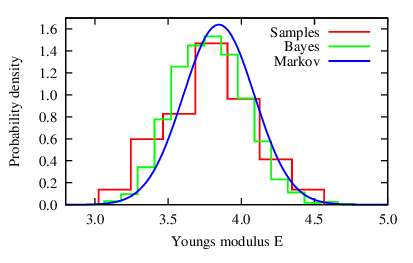}\\

\vspace{-2mm}
\includegraphics[width=0.43\textwidth]{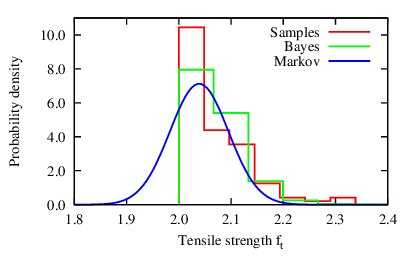}
\includegraphics[width=0.43\textwidth]{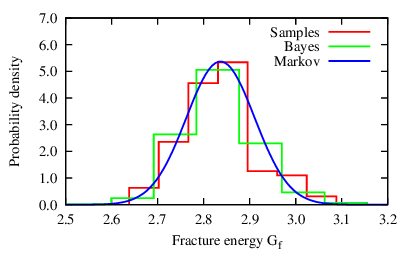}\\

\vspace{-2mm}
\includegraphics[width=0.43\textwidth]{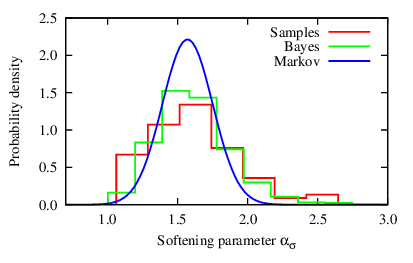}
\includegraphics[width=0.43\textwidth]{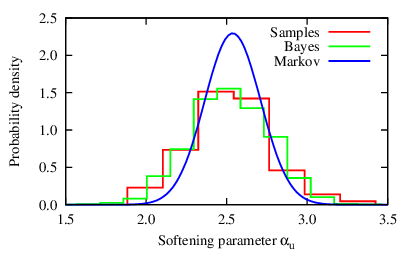}
\caption{Estimated parameter distributions for the wedge splitting test}
\label{wedge_histo}
\end{figure}

In the next step the parameter uncertainties are estimated with the different approaches by assuming a constant measurement noise of 
$\sigma_\epsilon = 0.5 \cdot 10^{6} \text{N/m}^2$.
In Table \ref{wedge_samples_results} and Figure \ref{wedge_histo} the results of these analyses are shown.
They indicate are very good agreement of the Bayes' estimates with the variation from the sample analysis.
The Markov estimator gives a good approximation of the variation, but some parameter dependencies are
estimated not correct.

\begin{table}[th]
\hspace{2mm}\begin{minipage}[c]{0.48\textwidth}
\centering
\small
\begin{tabular}{ccccccc}
    \hline
	&\multicolumn{2}{c}{Samples}&\multicolumn{2}{c}{Markov}& \multicolumn{2}{c}{Bayes}\rule[0mm]{0pt}{2.5ex}\\
	&  Mean & Std. & Mean & Std.& Mean & Std.\rule[-1mm]{0pt}{2ex}\\
	
   \hline
    $E$   		  	&3.78 & 0.30  &  3.85 & 0.24 & 3.79 & 0.24 \rule[0mm]{0pt}{2.5ex}  \\
	$f_t$ 		  	&2.07 & 0.07  &  2.04 & 0.06 & 2.07 & 0.05 \\
	$G_f$ 		  	&2.84 & 0.08  &  2.84 & 0.07 & 2.83 & 0.08 \\
	$\alpha_\sigma$ &1.62 & 0.32  &  1.57 & 0.18 & 1.62 & 0.26 \\
	$\alpha_u$ 	  	&2.51 & 0.26  &  2.54 & 0.17 & 2.52 & 0.24 \\
																								   
   \hline
  \end{tabular}

\end{minipage}
\begin{minipage}[c]{0.48\textwidth}
\begin{equation*}
\begin{aligned}
Corr_{\mathbf{pp}}^{Samples} &= \left[
\begin{array}{rrrrr} 
1.00 & \text{-}0.65 & 0.29 & \text{-}0.26 & 0.16\\
 & 1.00 & \text{-}0.13 & 0.40 & \text{-}0.34\\
 && 1.00 & \text{-}0.31 & \text{-}0.65\\
 &&& 1.00 & 0.41\\
 &&&& 1.00 \\
 \end{array}\right]
\end{aligned}
\end{equation*}
\end{minipage}\\

\hspace{2mm}\begin{minipage}[c]{0.50\textwidth}
\begin{equation*}
\begin{aligned}
Corr_{\mathbf{pp}}^{Markov} &= \left[
\begin{array}{rrrrr} 
1.00 & \text{-}0.59 & 0.29 & \text{-}0.83 & \text{-}0.40\\
 & 1.00 & \text{-}0.05 & 0.57 & \text{-}0.09\\
 && 1.00 & 0.02 & \text{-}0.79\\
 &&& 1.00 & 0.00\\
 &&&& 1.00 \\
 \end{array}\right]
\end{aligned}
\end{equation*}
\end{minipage}
\begin{minipage}[c]{0.47\textwidth}
\begin{equation*}
\begin{aligned}
Corr_{\mathbf{pp}}^{Bayes} &= \left[
\begin{array}{rrrrr} 
1.00 & \text{-}0.56 & 0.17 & \text{-}0.09 & 0.12\\
 & 1.00 & \text{-}0.02 & 0.21 & \text{-}0.24\\
 && 1.00 & \text{-}0.21 & \text{-}0.69\\
 &&& 1.00 & 0.49\\
 &&&& 1.00 \\
 \end{array}\right]
\end{aligned}
\end{equation*}
\end{minipage}
\caption{Parameter estimates and correlations for the wedge splitting test}
\label{wedge_samples_results}
\end{table}

\section{CONCLUSIONS}
The approximation of parameter uncertainties and dependencies by Bayes' updating is more accurate
then using the Markov estimator.
However, the Markov estimator is suitable as a rough estimate even for non-convex problems.
A remarkable reduction of the numerical effort is possible by using meta-model approach to represent the likelihood function.

\section{ACKNOWLEDGMENT}
This research has been supported by the German Research Council (DFG) through Research Training Group 1462, 
which is gratefully acknowledged by the author.


\end{document}